\documentclass[floatfix,aps,prb,twocolumn,superscriptaddress]{revtex4-2}

\usepackage{amsmath,amssymb,bm}
\usepackage{graphicx}
\usepackage{xcolor}
\usepackage{braket}
\usepackage{dcolumn}
\usepackage{mleftright}
\usepackage{xparse}
\usepackage[normalem]{ulem}
\usepackage{soul}
\usepackage{adjustbox}
\usepackage{tikz}
\usetikzlibrary{calc,positioning}
\usepackage{microtype}

\usepackage{hyperref}

\newcommand{\comment}[1]{}

\begin{document}
\title{Finite temperature precursors of Mottness in the Fermi Hubbard model} 
%\\
%Comparison between simulations and optical lattice emulators}
%Entropy and Moment Correlations in 2D Fermi-Hubbard model}



%\thanks{}
\author{Sayantan Roy}
\affiliation{Department of Physics, The Ohio State University, Columbus OH 43210, USA}
\affiliation{Arnold Sommerfeld Center for Theoretical Physics and Center for NanoScience,
Ludwig-Maximilians-Universit\"at M\"unchen, D-80333 Munich, Germany}
\affiliation{Munich Center for Quantum Science and Technology (MCQST),
Schellingstr. 4, M\"unchen D-80799, Germany}
\author{Abhisek Samanta}
\affiliation{Department of Physics, The Ohio State University, Columbus OH 43210, USA}
\affiliation{Department of Physics, Indian Institute of Technology
Gandhinagar, Palaj, Gujarat 382355, India}
\author{Nandini Trivedi}
\affiliation{Department of Physics, The Ohio State University, Columbus OH 43210, USA}


%%%%%%%%%%%%% %%%%%%%%%%%%%%%%%% TODO: %%%%%%%%%%%%%%%%%%%%%%%%%%%%%%%%%
% Latest batch of changes that should finish up my (Sameed's) part:
% Plan: finish by Friday, Jan 28
% - Plot Structure Factors
%     [ ] Heatmap below half filling for multiple temperatures
%     [ ] Heatmap at half filling for multiple temperatures
%     [ ] Heatmap above half filling for multiple temperatures
%     [ ] AF cf vs density and temperature
%     [ ] Ferro cf vs density and temperature
% - Replot graphs with hot/cold color scheme and/or no grid lines
%     [x] Cs/Cm vs distance
%     [ ] Entropy
%     [ ] Decomposition plots
%     [ ] Structure Factors
%%%%%%%%%%%%%%%%%%%%%%%%%%%%%%%%%%%%%%%%%%%%%%%%%%%%%%%%%%%%%%%%%%%%%%

\date{\today}

\begin{abstract}
%Cold atom systems provide a rich platform to realize strongly interacting condensed matter systems, and recent progress in fluorescence imaging technique has enabled identification of nontrivial doublon, singlon, and holon correlation functions. We report a determinantal quantum Monte Carlo (DQMC) study of such correlation functions in the two-dimensional repulsive Fermi Hubbard model on a square lattice as a function of doping, interaction strength and temperature. Our aim is to identify signatures of the crossover from small $U$ (band regime) to large $U$(correlated insulator regime). Our key findings are: (1) Opening of a charge gap through the temperature variation of the equation of state $n$ vs $\mu$ and correlations with the behavior of the temperature-dependent doublon number.
\noindent
We investigate finite-temperature precursors of Mottness in the repulsive Hubbard model above the spin-ordering temperature, using numerically exact determinant quantum Monte Carlo. We show that the finite-temperature crossover is accompanied by a pronounced suppression of charge fluctuations, despite the presence of a gapless single-particle spectra, demonstrating that Mottness first emerges via two-particle response in an anomalous metallic regime, before appearing in single-particle spectral functions. We further show that a gap formation in the density of states occurs through momentum-resolved redistribution of spectral weight across the Brillouin zone, that begins at the onset of anomalous metallic regime, rather than through gap formation in single-particle spectral functions at individual momenta. Upon doping, the anomalous-metallic regime generates  transport and spectroscopic signatures characteristic of doped Mott insulators. This shows that the transport and spectroscopic anomalies of the doped Hubbard model do not require the presence of a fully formed Mott insulator at half-filling, but instead originate from the strong charge-response renormalization and spectral-weight redistribution that develop within the precursor anomalous metallic regime at half-filling.
  %Our analysis is based on the behavior of holon and doublon correlators and insights gained from a low-energy parton description.
  
  %The The Seebeck coefficient anomaly doesn't vanish monotonically with respect to temperature, anomalous sign change is , and shows strong $U/t$ dependence,  The doping at which Seebeck coefficient changes sign is also the doping at which the thermodynamic entropy is maximized, and at which the system goes from an insulating to metallic behavior as dictated by the doublon number, $\frac{\partial d}{\partial T}$.} (3) \rcol{A parton description of the effective low energy hamiltonian with only charge degrees of freedom is sufficient to capture the anomalous sign change in Seebeck coefficient, showing role of charge gap in sign change of $S_{kelvin}$.}
\end{abstract}

%\keywords{}

\maketitle
%\section{Introduction}
%\subsection{}
%\subsubsection{}

%\bcol{Thermopower or the Seebeck coefficient is an important transport quantity which measures the efficiency of direct conversion from thermal to electrical energy. Moreover, it tracks the nature of the majority carriers in the system. In weakly interacting systems, where Fermi liquid theory holds, the Seebeck coefficient is positive when the excitations are electron-like, but changes to negative when excitations are hole-like~\cite{behnia2015fundamentals}.
%5
%However, the presence of strong correlations in the system, as observed in a large class of cuprates, lead to various \textit{anomalous} behavior, i.e. the abrupt change in sign and magnitude of the Seebeck coefficient. Such sign changes in the Seebeck coefficient can be attributed to the Fermi surface reconstruction and possible strange metal phases~\cite{gourgout2022seebeck}.}

\noindent
\textit{Introduction:}
The repulsive Hubbard model is a paradigmatic model for understanding correlation-driven metal-insulator transitions in strongly correlated systems~\cite{anderson1959new,mott1961transition,hubbard1963electron}. While weak interactions favor a Slater-type insulating state driven by magnetic instabilities of the Fermi surface~\cite{slater1951magnetic,schrieffer1989dynamic}, strong interactions localize electrons into a Mott insulator with emergent magnetic moments \cite{mott1961transition,imada1998metal,macdonald1988t}. The intermediate-coupling regime, particularly relevant to cuprate superconductors, remains far less understood because the many-body state evades controlled descriptions based on either weak or strong coupling limits. Understanding how Mottness emerges in this regime thus remains a central challenge in correlated materials.

The situation becomes particularly challenging at finite temperatures, where competing thermal, charge, and spin fluctuations generate extended crossover regimes and pseudogap phenomena~\cite{schafer2021tracking}. Earlier studies have  tackled the role of quasi-long range antiferromagnetic (AFM) fluctuations in driving an extended metal-insulator crossover at low temperatures~\cite{schafer2021tracking,kim2020spin,lenihan2021entropy,vsimkovic2020extended}, where spin fluctuations lead to a momentum-differentiated onset of gap in the single particle spectra. However, such correlations necessitates one to remain below the ordering temperature for the spin moments in finite sized systems; hence it is pertinent to ask how does the metal-insulator crossover scenario evolves at higher temperatures, $T>T_{\rm spin}$~\cite{paiva2010fermions}.

The higher temperature regime is also particularly interesting in the light of development of optical lattice based quantum emulators over the last few years~\cite{bloch2005ultracold,bloch2012quantum,bakr2009quantum,aidelsburger2013realization,gross2017quantum}, which predominantly operate at intermediate temperature, and have revealed a wealth of information on different phases of the Hubbard model~\cite{koepsell2019imaging,koepsell2020robust,hirthe2023magnetically,salomon2019direct,vijayan2020time,koepsell2021microscopic,sompet2022realizing,chiu2018quantum,chiu2018quantum,chiu2019string}. This necessitates  to have a description of the metal-insulator crossover scenario in such a temperature regime, where experiments can directly probe the microscopic correlations across the crossover.

To connect finite-temperature Hubbard physics to anomalous metallic behavior in correlated materials, we expand on an extended metal-insulator crossover occurring above $T_{\rm spin}$, reported by some of us~\cite{roy2025metal}. The crossover originates from a separation between the interaction strengths at which the compressibility and single-particle density of states become insulating-like, and is amenable to direct probing in cold-atom experiments. Previous DMFT studies have shown that nonperturbative precursors of the Mott transition can manifest through divergent two-particle vertex functions well inside the metallic regime~\cite{schafer2013divergent,pelz2023highly}. More recent numerical studies have further identified an extended bad-metallic and pseudogap-metal regimes, together with anomalous incoherent transport, across a finite-temperature crossover in the Hubbard model~\cite{lu2026quantum,lu2026thermal,eom2025strange}. However, several key questions remain: What is the microscopic origin of this anomalous crossover, beyond purely local descriptions? How does spectral weight reorganize across the Brillouin zone as the system crosses over from the metallic to the insulating regime? Does Mottness first emerge in two-particle response functions before appearing in single-particle spectra? And more broadly, do the anomalous transport and spectroscopic signatures of doped Mott systems originate from a fully developed Mott insulator, or from finite-temperature precursor Mottness emerging already within the crossover regime~\cite{comanac2008optical}? %And how are the anomalous transport and spectroscopic signatures of doped Mott insulators connected to this finite-temperature crossover? In addition, whether the transport and spectroscopic anomalies of doped Mott systems originate from a fully developed Mott insulator or from finite-temperature precursor Mottness remains unresolved~\cite{comanac2008optical}.

In this letter, we address these questions by identifying precursor Mottness as the organizing principle underlying the extended crossover, characterized by the separation between insulating behavior in compressibility and in single-particle spectra, strong charge response renormalization despite a gapless single-particle spectra, and momentum-resolved redistribution of spectral weight preceding a pseudogap formation in the density of states. We examine the behavior of two particle response functions that mediate the electronic compressibility, spectral functions over the Brillouin zone, and thermoelectric transport coefficient $S_{\rm Kelvin} = -\frac{\partial s}{\partial n}$, due to its natural connection with thermodynamic variables~\cite{shastry2013thermopower,peterson2010kelvin}. We find that the charge response begins to be increasingly suppressed at the onset of an anomalous-metallic (AM) regime, although the single-particle density of states shows no gap formation. At finite temperature, onset of Mottness first occurs in two-particle responses, rather than single-particle spectra. The crossover is also accompanied by redistribution of low energy spectral weight in the Brillouin zone, which starts at the onset of the AM regime, and ultimately leads to the formation of a gap in $N(\omega)$, giving rise to the insulating regime. In addition, doping the AM regime at half filling (that shows absence of a fully formed Mott gap) leads to an anomalous metallic state with a reconstructed Fermi surface and anomalous sign of the Seebeck coefficient near half filling. Thus, the anomalous metallic regime serves as the precursor phase from which transport and spectroscopic anomalies, characteristic of the doped Mott insulator, originate.

\begin{figure}{\includegraphics[width=0.8\linewidth]{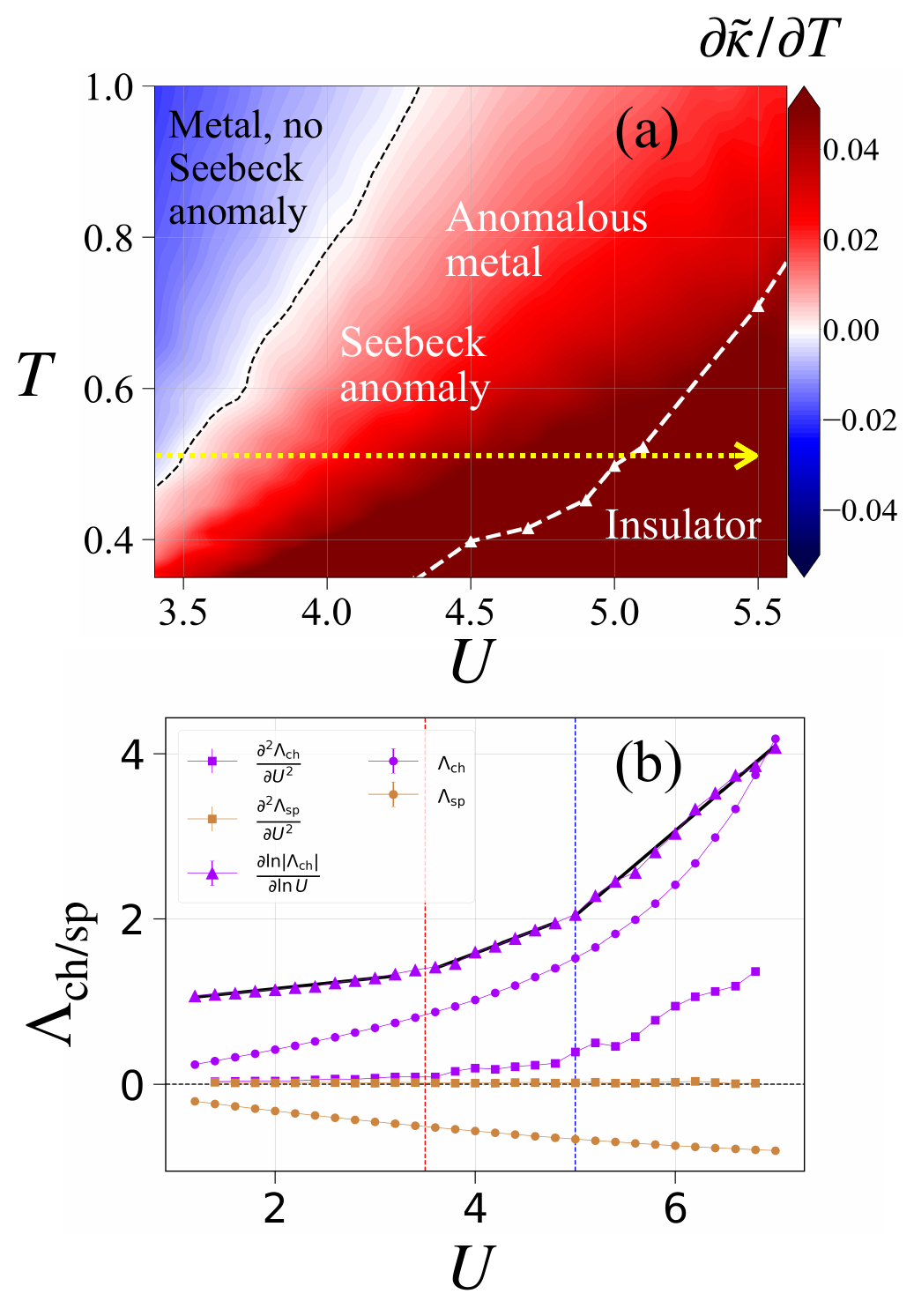}}
\caption{Anomalous metal-insulator crossover in the repulsive Hubbard model at half-filling. \textbf{(a)} Crossover diagram in the $T-U$ plane showing metallic, anomalous-metallic (AM), and insulating regimes. The colormap represents the sign of $\partial \tilde{\kappa}/\partial T$,  where  $\tilde{\kappa} =\partial n/\partial \mu$ (TDOS); positive (negative) values indicate insulating-like (metallic-like) behavior~\cite{kim2020spin,roy2025metal}. The two crossover lines correspond to the onset of insulating behavior in TDOS and in $N(\omega)$ (DOS), respectively (see text). Doping the AM and insulating regimes generates an anomalous sign change of the Seebeck coefficient near half filling. \textbf{(b)} Charge and spin response renormalization parameters $\Lambda_{\rm ch/sp}$ at $T=0.5$(along yellow arrow in panel (a)). Red and blue dashed lines denote $U_{\rm cr}^{\rm TDOS}(T)$ and $U_{\rm cr}^{\rm DOS}(T)$ respectively. Beyond $U_{\rm cr}^{\rm TDOS}(T)$, $\Lambda_{\rm ch}$ start to increase rapidly (shown by $\partial \Lambda_{\rm ch}^2/\partial U^2$ and $\partial \ln |\Lambda_{\rm ch}|/\partial \ln U$) while the corresponding spin-channel quantities remain weakly affected by the crossover. The first signature of precursor Mottness therefore appears in the AM regime.} 
\label{Figure1: Intro}
\end{figure}

\medskip
\noindent
\textit{Model and method:} We consider the single band Fermi Hubbard model on a square lattice with nearest neighbor hopping and onsite repulsive interaction,
$\mathcal{H}= -t\sum_{\langle ij \rangle,\sigma}{  \hat{c}^\dag_{i\sigma} \hat{c}_{j\sigma} }-\mu \sum_{i}\hat{n}_{i}+U\sum_{i}{ \left( \hat{n}_{i\uparrow} - \frac{1}{2} \right)\left( \hat{n}_{i\downarrow} - \frac{1}{2} \right)}
$. The operators
 $\hat{c}_{i\sigma}$ and $\hat{c}^\dag_{i\sigma}$ are fermionic annihilation and creation operators, respectively.
The number operator is defined as $\hat{n}_{i,\sigma} \equiv \hat{c}^\dag_{i\sigma} \hat{c}_{i\sigma}$,  $\hat{n}_i = \hat{n}_{i \uparrow} + \hat{n}_{i\downarrow}$, and the particle density per site ${n} = \sum_i{\langle\hat{n}_i\rangle}/{N_s}$, where $N_s$ is the total number of sites. We define hopping amplitude $t$ as the unit of energy, $\mu$ is the chemical potential and $U$ is the onsite Coulomb repulsion. We perform numerically exact Determinantal Quantum Monte Carlo (DQMC) simulations~\cite{blankenbecler1981monte,hirsch1985two} at intermediate to high interaction strengths, that treats both local and non-local correlations exactly. 
%We also restrict ourselves to intermediate to high temperature, where the Kelvin formula is justified, and sign problem is not an issue.
We further perform analytic continuation using the maximum entropy package CQMP-MaxEnt~\cite{levy2017implementation}, with default models chosen to optimize the sum rules~\cite{white1991spectral,swanson2014dynamical}.

\medskip
\noindent
\textit{Anomalous metal to insulator crossover at half-filling}: We begin by reviewing two different metrics to distinguish between metals and insulators. A metallic system has a finite compressibility (thermodynamic density of states, TDOS), whereas an insulator is incompressible at $T=0$. At finite temperatures, an insulator can exhibit finite compressibility due to thermally activated density fluctuations~\cite{kim2020spin,roy2025metal}; hence sign of the temperature derivative of TDOS can distinguish between a metal and an insulator (Fig.~\ref{Figure1: Intro}(a)). In addition, the single-particle density of states (DOS) $N(\omega)=\frac{1}{N}\sum_{k}A(k,\omega)$ for a metal has a peak at $\omega=0$. In contrast, an insulator exhibits a gap in DOS at the Fermi energy. In Landau Fermi liquid theory, these two metrics are related by the dimensionless symmetric Landau parameter, $\partial n/\partial \mu = N(0)/(1+F^{s}_{0})$. Hence, a crossover from a ``Fermi liquid" metal to an insulator should preserve this relation. The onset of a gap in the TDOS should be accompanied by a corresponding gap in the DOS. However, in the single band repulsive Hubbard model, this criterion is violated~\cite{roy2025metal}. At fixed temperature, the TDOS exhibits a gap opening at a crossover strength $U_{\rm cr}^{\rm TDOS}(T)$, while the DOS develops a gap at a larger value $U_{\rm cr}^{\rm DOS}(T)>U_{\rm cr}^{\rm TDOS}(T)$. Hence, the half filled Hubbard model exhibits an anomalous metal to insulator crossover, with 3 distinct regimes shown in Fig.~\ref{Figure1: Intro}(a): (i) A metallic regime for $U<U_{\rm cr}^{\rm TDOS}(T)$ (ii) A transitional \textit{anomalous-metallic} (AM) regime  for $U_{\rm cr}^{\rm TDOS}(T)<U<U_{\rm cr}^{DOS}(T)$ (shown between the black and white dashed lines in Fig.~\ref{Figure1: Intro}(a)) which exhibits a gap in TDOS, but no gap in DOS (iii) Insulating regime for $U>U_{\rm cr}^{\rm DOS}(T)$ which is gapped in both DOS and TDOS. In the following, we use the term ‘insulating regime’ to denote the onset of a gap in the single-particle density of states, while reserving the term ‘Mott insulator’ for the case where $N(\omega)$ develops a hard gap, and shows well separated Hubbard bands at large-$U$.

\begin{figure}{\includegraphics[width=\linewidth]{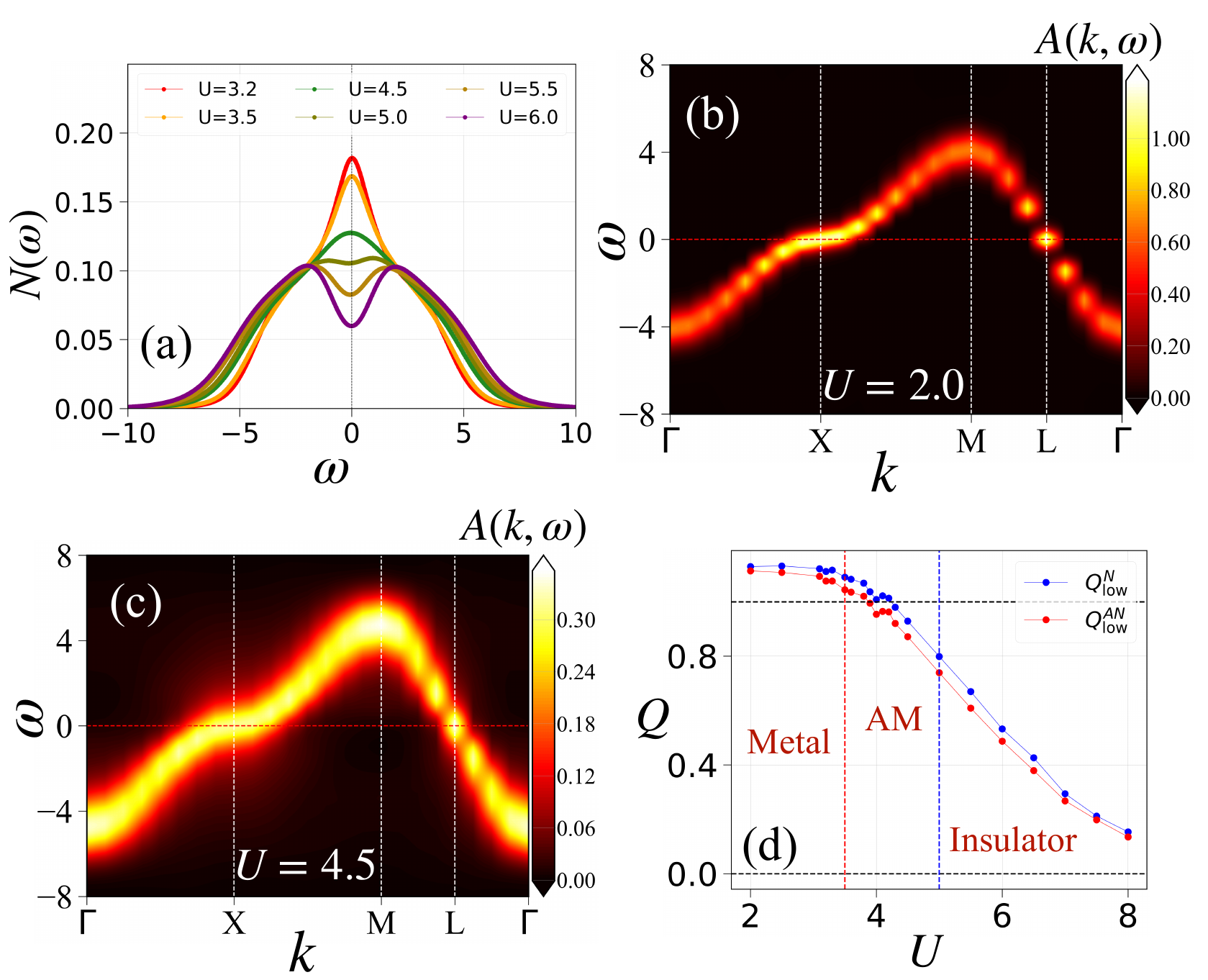}}
	\caption{Single-particle spectral functions across the extended metal-insulator crossover at half filling, $T=0.5$. \textbf{(a)} As a function of $U$, the $N(\omega)$ evolves from having a maxima at $\omega=0$ to formation of a dip around $\omega = 0$, for $U \geq 5.0$. \textbf{(b)-(c)} Evolution of single-particle dispersion $A(k,\omega)$ along a high symmetry path in the brillouin zone. As the system crossovers from a metallic regime (panel \textbf{(b)}) to the anomalous metallic regime (panel{(\textbf{c})}), spectral weight shift from the $X$ and $L$ points to $M$ and $\Gamma$ point.  \textbf{(d)} Spectral weight ratio $Q$ (defined in main text) across the crossover, shown both for $X$ and $L$ points. Red dashed line marks $U_{\rm cr}^{\rm TDOS}$, blue dashed line marks $U_{\rm cr}^{\rm DOS}$. The onset of spectral weight transfer occurs at metal-AM crossover boundary.} 
	\label{Figure2: Dos_gap}
\end{figure}

We look at the two-particle charge response, responsible for the insulating like behavior in TDOS. The TDOS corresponds to the static limit of the response~\cite{peterson2010kelvin};
\begin{align}
	\frac{\partial n}{\partial \mu} = \chi_{\rm ch}(q=0,\omega_n=0) &= \frac{1}{\beta}\int_{0}^{\beta}d\tau \langle \hat{n}(0,\tau)\hat{n}(0,0)\rangle
\end{align}
We define the renormalization of the static charge response as $\chi_{\rm ch} = \chi_0/(1+\Lambda_{\rm ch})$, where $\chi_{0}$ is the particle-hole bubble~\cite{supplemental}. For Fermi liquid, $\Lambda_{\rm ch}$ corresponds to the symmetrized Landau parameter, $F_{0}^{s}$, and $\chi_{0} = N(0)$, the 1-particle density of states at the Fermi level. We plot the renormalization parameter $\Lambda_{\rm ch}$ vs $U$ at half filling in Fig.~\ref{Figure1: Intro}(b). For $U<U_{\rm cr}^{\rm TDOS}(T)$, the charge response is weakly renormalized. Once the system enters the AM regime, $\Lambda_{\rm ch}$ grows rapidly with increasing $U$, corroborated by both $\partial \ln|\Lambda_{\rm ch}|/\partial \ln U$ and $\partial^2\Lambda_{\rm ch}/\partial U^2$. This trend continues into the insulating regime. In contrast, the corresponding spin-response renormalization shows no comparable change across the crossover. The charge sector therefore begins to depart from weak-coupling behavior already at $U_{\rm cr}^{\rm TDOS}(T)$,  although the single-particle density of states remains gapless. The separation between these two crossover scales demonstrates that the onset of Mottness is first reflected in two-particle response functions before appearing in the single-particle spectra. We also look at the renormalization of the spin response, $\chi_{\rm sp} = \chi_{0}/(1+\Lambda_{\rm sp})$; and find that it exhibits no distinct behavior across the crossover. Hence, this anomalous crossover is accompanied by a significant enhancement of vertex contributions to the charge response and onset of  localization of the charge degrees of freedom, although the spin response is weakly renormalized. 

%This is expected, since the crossover occurs at $T>T_{\rm spin}$, the spin ordering temperature. 

\medskip
\noindent
\textit{Redistribution of spectral weight across the extended crossover}: We now turn to single-particle response functions across the extended crossover. In the AM regime, although the charge degrees of freedom start getting pinned down, the system exhibits ``resilient" itinerant single-particle excitations, in form of a peak in $N(\omega)$ at $\omega \sim 0$ for $U<U_{\rm cr}^{\rm DOS}(T)$, shown in Fig.~\ref{Figure2: Dos_gap}(a). To probe the single-particle dispersion $A(k,\omega)$ across the crossover, we look along a high symmetry path in the Brillouin zone from $\Gamma (0,0) \rightarrow X(0,\pi) \rightarrow M(\pi,\pi) \rightarrow L(\pi/2,\pi/2) \rightarrow \Gamma$. For a metal at half filling, most of the spectral weight of $A(k,\omega)$ occurs around the antinode $X$ and the node $L$ points. In contrast, for a Mott insulator, most of the spectral weight occurs around $\Gamma$ (lower Hubbard band) and $M$ (upper Hubbard band) points. The extended crossover bridges between these two scenario~\cite{supplemental}, with onset of spectral weight transfer between $X,L$ to $\Gamma,M$ points as the system crossovers from the metal to the AM regime (shown in Fig.~\ref{Figure2: Dos_gap}(b-c).

To quantify the spectral weight transfer, we look at thermal spectral weight $N_{T}(k)$ at the high symmetry points
\begin{align}
   N_{T}^{K} &=\int_{-\infty}^{\infty}d\omega \left(-\frac{\partial f_K}{\partial \omega}\right)A(K,\omega)
\end{align}  
where $f$ is the Fermi distribution function at temperature $T$; $f_{X/L}$ is centered at $\omega=0$ for the nodal/antinodal points, and $f_{\Gamma/M}$ is centered at $\omega_{\Gamma/M}$, (location of peak spectral intensity at the $\Gamma/M$ points). Note that due to the derivative $-\partial f/\partial \omega$, this quantity picks up the relevant low energy spectral weight in the interval $[-T,T]$ relative to the center of $f$, and accounts for smearing of $A(k,\omega)$ at finite temperatures. We also define $Q^{N}_{\rm low} = N_{T}^{L}/N^{\Gamma}_{T}$, and $Q^{AN}_{\rm low} = N_{T}^{X}/N_{T}^{\Gamma}$, which counts the spectral weight at $X/L$ point relative to the $\Gamma$ point (shown in Fig.~\ref{Figure2: Dos_gap}(h)). On crossover to the AM regime at $U_{\rm cr}^{\rm TDOS}(T)$, the spectral weight starts to redistribute from the $X/L$ points, captured by a decline in $Q^{\rm N/AN}_{\rm low}$. The AM regime is remarkable for the following reason: Even though resilient low energy excitations (in the form of a gapless $N(\omega)$) exist despite enhanced suppression of charge fluctuations, (see Fig.~\ref{Figure1: Intro}(b)), the spectral weight responds to the onset of Mottness in the two-particle response level, via this redistribution. At $U_{\rm cr}^{\rm DOS}(T)$, $N(\omega)$ shows a gap onset around $\omega=0$ due to this redistribution, and the individual spectral functions show gap onset at larger $U$. Thus, at finite temperature, the insulating regime at $U_{\rm cr}^{\rm DOS}(T)$ is realized not by a gap formation in the individual $A(k,\omega)$, but rather by a redistribution of spectral weight away from the $k$ points on the Fermi surface~\cite{supplemental}. 

\begin{figure}{\includegraphics[width=\linewidth]{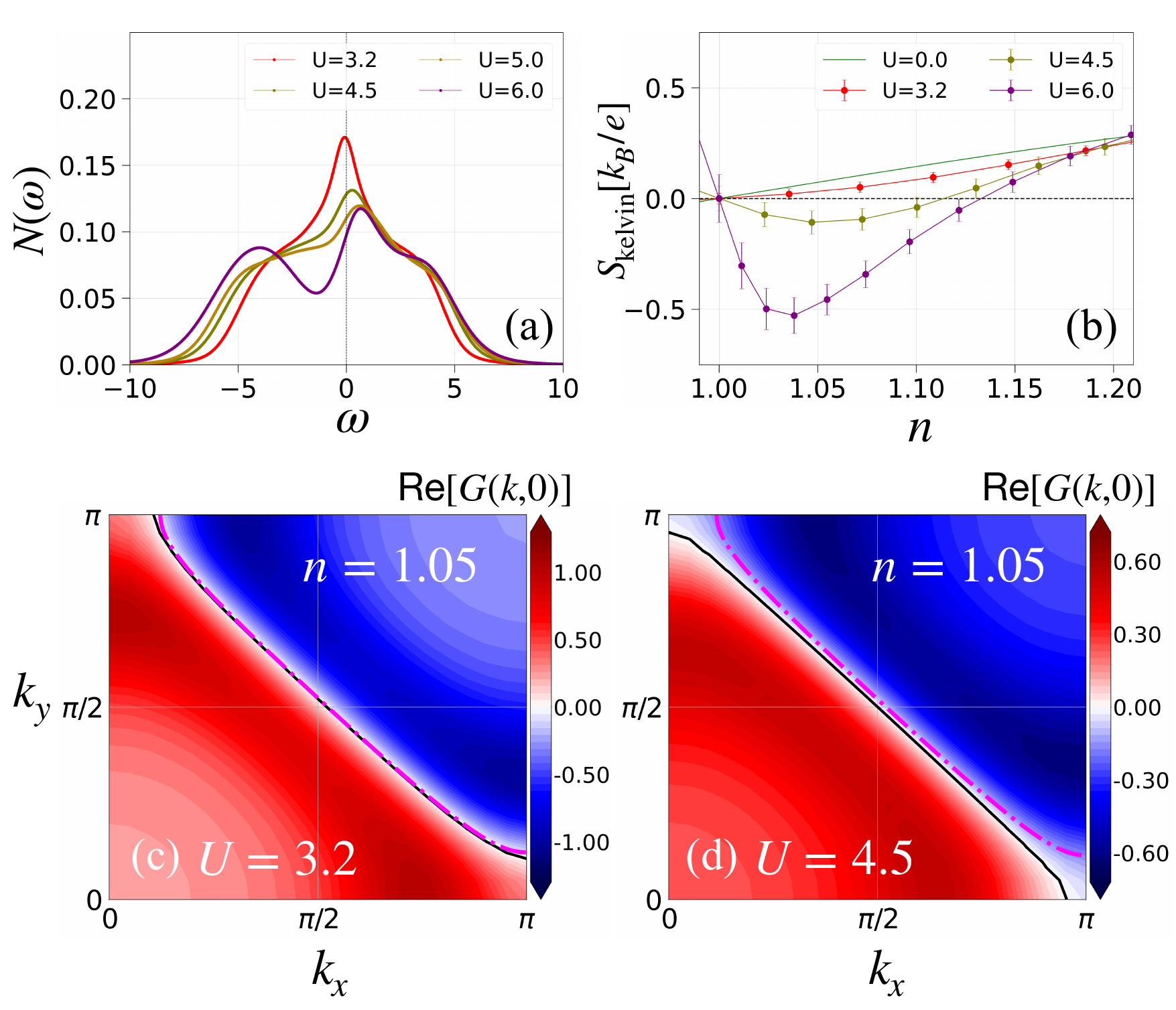}}
	\caption{Spectroscopic and transport signatures upon doping the three finite-temperature crossover regimes at $T=0.5$, along the dashed yellow arrow in Fig.~\ref{Figure1: Intro} \textbf{(a)} $N(\omega)$ at $n=1.05$. Doping the metallic system at half filling ($U=3.2$) results in a single peak around $\omega=0$. However, doping the system from AM phase ($U=4.5$) results in development of a shoulder at negative $\omega$. When the system is doped away from the insulator ($U=6.0$), $N(\omega)$ exhibits a 3-peak structure. \textbf{(b)} Seebeck coefficient vs doping for different $U$. The anomalous sign change of the Seebeck coefficient first appears when the system is doped from the AM regime at half filling (shown for $U = 4.5$). \textbf{(c)} Fermi surface (black solid line) for the doped metallic system ($U=3.2$) encloses a large volume, and is consistent with sign of the Seebeck coefficient. \textbf{(d)} Fermi surface of the doped AM system ($U=4.5$) encloses a smaller volume. However, the topology is consistent with the sign of Seebeck coefficient. The magenta dashed line marks the Fermi surface for a non-interacting system with the same density.} 
	\label{Figure3: Dos_Fermi_surface}
\end{figure}

\medskip
\noindent
\textit{Spectroscopic and transport anomalies at finite doping}: For doped Mott Insulators, $N(\omega)$ exhibits an anomalous structure: a low energy peak around $\omega=0$, and remnants of an upper and lower Hubbard bands, ~\cite{dagotto1991density,osborne2021broken}. To study the origin of this behavior, we reexamine the behavior of $N(\omega)$ as the system is doped from various regimes at half filling in Fig~\ref{Figure1: Intro}(a): the metallic regime, the AM regime and the insulating regime (shown in Fig.~\ref{Figure3: Dos_Fermi_surface}). For $ U_{\rm cr}^{\rm TDOS}(T)<U<U_{\rm cr}^{\rm DOS}(T)$, $N(\omega)$ for low particle dopings ($n=1.05$) shows a peak near $\omega=0$.

A closer look at $N(\omega)$ reveals that anomalous spectral features start to emerge already from the AM regime. Doping the metallic regime ($U<U_{\rm cr}^{\rm TDOS}(T)$) produces a single low-energy peak near $\omega=0$. In contrast, doping the insulating regime ($U_{\rm cr}^{\rm DOS}(T)$) yields the characteristic three-peak structure of a doped Mott insulator, consisting of a lower Hubbard band at $\omega<0$, itinerant low-energy excitations near $\omega=0$, and a remnant upper Hubbard band at $\omega>0$. Remarkably, a lower Hubbard band like spectral feature already begins to develop upon doping the AM regime, where $N(\omega)$ acquires an asymmetric broad shoulder at $\omega<0$ that evolves continuously with $U$ into a well-separated lower Hubbard band for $U>U_{\rm cr}^{\rm DOS}(T)$. This is unexpected, because the half-filled AM regime itself does not exhibit incipient Hubbard bands. This suggests that for the doped AM regime, $k$ states near the Fermi level start to shift spectral weight to lower energies $\omega<0$, (with the primary contributions from states near $(0,\pi)$~\cite{supplemental}. This is analogous to the spectral weight redistribution occurring for half-filling shown in Fig~\ref{Figure2: Dos_gap}(d).

The redistribution of spectral weight from states near the Fermi surface naturally raises the question of how the low-energy excitations evolve upon doping. To probe this behavior, we examine thermoelectric transport across the three crossover regimes identified above. In the Kelvin approximation, the Seebeck coefficient is given by $S_{\rm kelvin} = -(\partial s/\partial n)_{T}$, where $s$ is the thermodynamic entropy and $n$ the particle density~\cite{peterson2010kelvin,deng2013bad}. At intermediate to high temperatures, this approximation provides a reliable description of thermopower in strongly correlated systems~\cite{wang2023quantitative,mravlje2016thermopower,deng2013bad}. A characteristic feature of doped Mott systems is the anomalous sign change of the Seebeck coefficient near half filling~\cite{gourgout2022seebeck,cyr2017anisotropy,mandal2019anomalous}, reflecting a change of nature of the low-energy carriers. We now examine how this anomaly develops across the finite-temperature crossover considered here.

Due to the Maxwell relation $(\partial s/\partial \mu)_{T}=(\partial n/\partial T)_{\mu}$, the shape of the entropy curves $s(\mu,T)$ directly determines the Seebeck coefficient in the Kelvin formula [Fig.~\ref{Figure3: Dos_Fermi_surface}(b)]. For $U<U_{\rm cr}^{\rm TDOS}(T)$, the sign of Seebeck coefficient as a function of doping is in agreement with Fermi liquid prediction. Upon entering the AM regime, the insulating-like charge response $\partial \tilde{\kappa}/{\partial T}>0$ at half filling causes the maxima of entropy to shift to finite doping~\cite{supplemental}, producing the anomalous sign change of Seebeck coefficient characteristic of doped Mott insulators~\cite{wang2023quantitative,gourgout2022seebeck,cyr2017anisotropy,mandal2019anomalous,doiron2013hall}. The onset of anomalous sign change of $S_{\rm kelvin}$ with increasing $U$ therefore coincides with the finite-temperature crossover into the AM regime, indicating that anomalous transport behavior seen for doped Mott insulators already emerges from the precursor AM regime at a lower $U$.

Since the change of sign of Seebeck coefficient indicates a change of nature of carriers, we examine low energy excitations as visible in photoemission spectroscopy, from plotting the Fermi surface of the doped system in Fig.~\ref{Figure3: Dos_Fermi_surface}(c-d). As the low energy spectral features are smeared at finite temperature, we extract the Fermi surface as contours of sign change of $\text{Re}G(k,\omega=0)$, instead of $\text{max} [A(k,\omega=0)]$. When the system is doped from metallic regime at half filling $(U=3.2)$, the interacting Fermi surface encloses a ``large" volume, and coincides with the Fermi surface of a non-interacting system at same density. In contrast, when the system is doped from the AM ($U=4.5$), the Fermi surface encloses a small volume, deviates from the non-interacting Fermi surface, and signals a change in carrier type consistent with the sign of $S_{\rm kelvin}$. %Thus, similar to the anomalous Seebeck coefficient, the Fermi surface reconstruction observed for doped Mott Insulators also begins from doping the AM regime~\cite{supplemental}, which precedes the Mott Insulator at half filling. 
Together with the anomalous Seebeck response, the Fermi surface reconstruction demonstrates that the finite-temperature AM regime acts as the precursor from which anomalous metallic behavior, as seen for doped Mott insulators at higher $U$~\cite{osborne2021broken,RoyPrep}, emerges upon doping.

\medskip
\noindent
\textit{Conclusion and outlook}: Through exact numerical calculations, we have shown that the finite-temperature metal-insulator crossover in the repulsive Hubbard model exhibits a precursor anomalous-metallic (AM) regime, where Mottness first emerges in two-particle response functions before appearing in single-particle spectra. The onset of the AM regime is marked by strongly suppressed charge fluctuations despite the persistence of gapless low-energy excitations, signaling a breakdown of the conventional Fermi-liquid relation between compressibility and single-particle density of states. We further demonstrated that a gap formation in the single-particle density of states is driven not by gap opening at individual momenta, but by momentum-resolved redistribution of spectral weight across the Brillouin zone, which onsets at the metal-AM crossover boundary. These results identify a microscopic mechanism for the finite-temperature crossover and establish the AM regime as the precursor regime where Mottness first emerges at half-filling.

In addition, the precursor AM regime already contains the origin of anomalous metallic behavior that is characteristic of doped Mott insulators. Upon particle doping, the AM regime starts to generate lower Hubbard band like spectral feature in $N(\omega)$, anomalous Seebeck response, and a reconstructed Fermi surface, even before the system enters the fully developed Mott-insulating regime at half-filling. Our results therefore provide a unified framework connecting spectral and thermodynamic (compressibility and entropy) anomalies, momentum-space spectral weight redistribution, and anomalous transport in strongly correlated systems.  More broadly, our results suggest that the unconventional metallic behavior of doped Mott systems originates not from the fully formed Mott insulator itself, but from a finite-temperature precursor Mottness emerging already within the AM regime. 
Together with the observation of Ref.~\cite{comanac2008optical} that cuprates may not be doped Mott insulators in the strong coupling regime, our results point toward an intermediate-coupling origin of doped Mott insulator-like transport and spectroscopic anomalies that does not rely on the existence of a fully developed Mott gap. %\textcolor{red}{To check: should we keep this connection to single site DMFT?}Recent single-site DMFT studies of the Hubbard model have identified divergent vertex contributions to two-particle response functions in the coexistence regime of the metal-insulator transition~\cite{pelz2023highly}. The precursor Mottness identified here suggests a possible finite-temperature continuation of such vertex-contribution driven behavior into the extended crossover regime. %, providing a potential bridge between DMFT vertex divergences and anomalous metallic behavior in the doped Hubbard model.

Our description of the extended crossover also allows one to probe the AM regime and the onset of precursor Mottness directly in cold-atom quantum simulators, due to the presence of equal time correlation functions that track this crossover~\cite{roy2025metal}; the anomalous charge correlations can be probed directly in experiments. It will also be interesting to examine how the crossover evolves in the presence of next-nearest-neighbor hopping $t'$ where tuning magnetic frustration and particle-hole asymmetry can qualitatively modify the low energy behavior~\cite{zhang2025frustration}.
% and particle-hole asymmetry can qualitatively modify the redistribution of spectral weight and the associated transport anomalies and Fermi surface reconstruction. 
Finally, the present work motivates a detailed investigation of the full frequency dependence of two-particle charge and spin responses across the crossover, particularly in light of recent studies relating anomalous transport and incoherent metallic behavior to pseudogap-metal regimes in the Hubbard model~\cite{eom2025strange}. An analysis of how anomalous two-particle responses directly influence the low energy structure of the self energy at specific $k$ points in the Brillouin zone is left as direction of future research.

\medskip
\noindent
\textit{Acknowledgments:} S.R. acknowledges very fruitful discussions with Mathias Pelz, Jan von Delft, Andre Marie Tremblay, Alessandro Toschi, and Mohit Randeria. S.R., A.S. and N.T. acknowledge support from NSF Materials Research Science and Engineering Center (MRSEC) Grant No. DMR-2011876 and NSF-DMR 2138905. A.S. acknowledges support from ANRF, DST, India (File No. ANRF/ECRG/2024/003287/PMS). Computations were performed at the Unity cluster of Arts and Science College, Ohio State University.

%\textbf{Main text: 3225 words, Captions: another 498 words. Total = 3722 words. Abstract+Acknowledgement: Does not contribute to word count: 220 words }

\clearpage

\clearpage

\begin{center}
{\bf End Matter}
\end{center}
\noindent
Section~\ref{FS reconstruction} of the end matter provides details on the Fermi surface reconstruction of the doped model, while section ~\ref{Spectral weight} discusses additional single-particle spectral details for the doped model.

\section{Fermi surface reconstruction for the doped anomalous metal}
\label{FS reconstruction}
\begin{figure}[h]
{\includegraphics[width=0.8\linewidth]{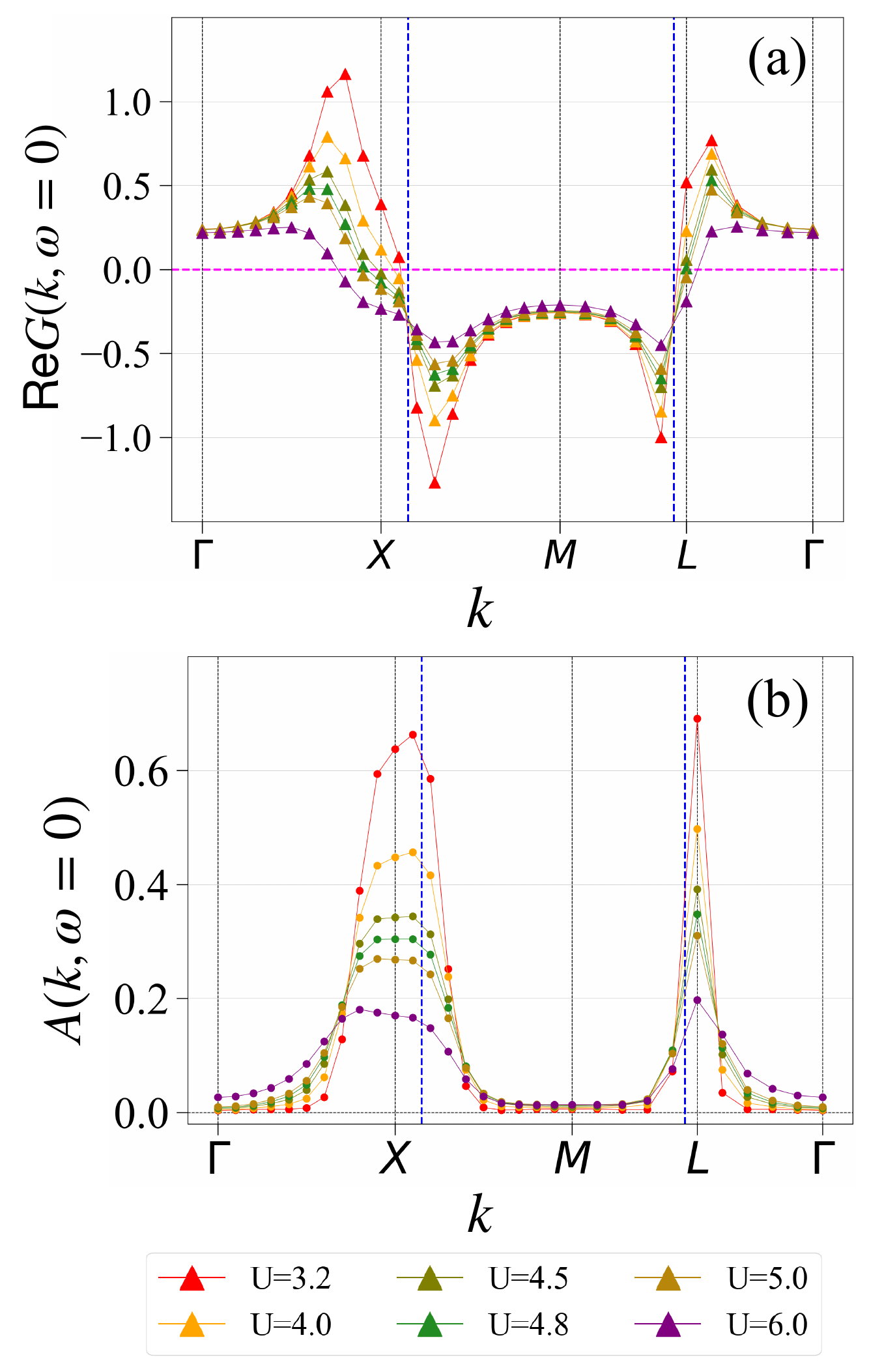}}
\caption{Single particle spectra at fixed density $n=1.05$, at $T=0.5$, along high symmetry path. \textbf{(a)} $\text{Re}G(k,\omega=0)$ along the high symmetry path. The sign change of $\text{Re}G(k,\omega=0)$ in $k$ space defines the location of the Fermi surface. The dashed vertical line marks the sign change for a non-interacting system at same density. \textbf{(b)} Momentum distribution curves $A(k,\omega=0)$ along the high symmetry path. As the system is doped from the anomalous metallic regime at half-filling $U=4.0, 4.5, 4.8$, it undergoes a Fermi surface reconstruction, most prominent near the $X=(0,\pi)$ point (see text).} 
\label{Figure1_end_matter}
\end{figure}
\noindent
In main text Fig.~\ref{Figure3: Dos_Fermi_surface}, we showed the contour of sign change of $\text{Re}G(k,\omega=0)$ (defining the Fermi surface) as the system is doped from the metallic, AM and insulating regime at half filling. Here, we examine the sign change for the largest lattice size in our calculation, $N=20\times 20$. Because of finite size of the lattice in QMC simulations, the location of the Fermi surface is determined upto an uncertainty $\delta k_{x} \times \delta k_y$, with $\delta k_{i} = 2\pi/N_{i}$ ($N_{i}$ being the number of lattice sites along $(i=x,y)$ direction). Hence, to rigorously define the Fermi surface reconstruction, we proceed as follows. The blue dashed line in Fig.~\ref{Figure1_end_matter} shows the location of sign change of $\text{Re}G(k,\omega=0)$ in $k$ space ($k_{NI}$) for a non-interacting system with the same density. This can be obtained upto a very high precision in $k$ space. For the interacting system, we define the following: let $(k_a,k_b)$ be adjacent $k$ points on the high symmetry path satisfying $\text{Re}G(k_a,\omega=0)>0$, and $\text{Re}G(k_b,\omega=0)<0$). (i) If $K_{NI} \in [k_a,k_b]$, the Fermi surface for the interacting system is identical to that of the non-interacting system within finite size resolution. (ii) If $k_{NI} \notin [k_a,k_b]$, then the Fermi surface for the interacting system is reconstructed with respect to the non-interacting Fermi surface.

The sign change of $\text{Re}G(k,\omega=0)$ along the high symmetry path is shown in Fig.~\ref{Figure1_end_matter}(a), for various $U$ values representing the three crossover regimes at half-filling. When the system is doped from the metallic regime at half filling, ($U=3.2$), the Fermi surface has the same topology and size as that of the non-interacting system. However, when the system is doped from the AM regime at half filling ($U=4.0, 4.5, 4.8$), the Fermi surface undergoes a reconstruction, with most prominent deviation happening near the $(0,\pi)$ point. For $U=4.5, 4.8$, the deviations near the $(0,\pi)$ point are strong enough to change the topology of the Fermi surface from ``large/concave" to ``small/convex".

\begin{figure*}[t]
{\includegraphics[width=\linewidth]{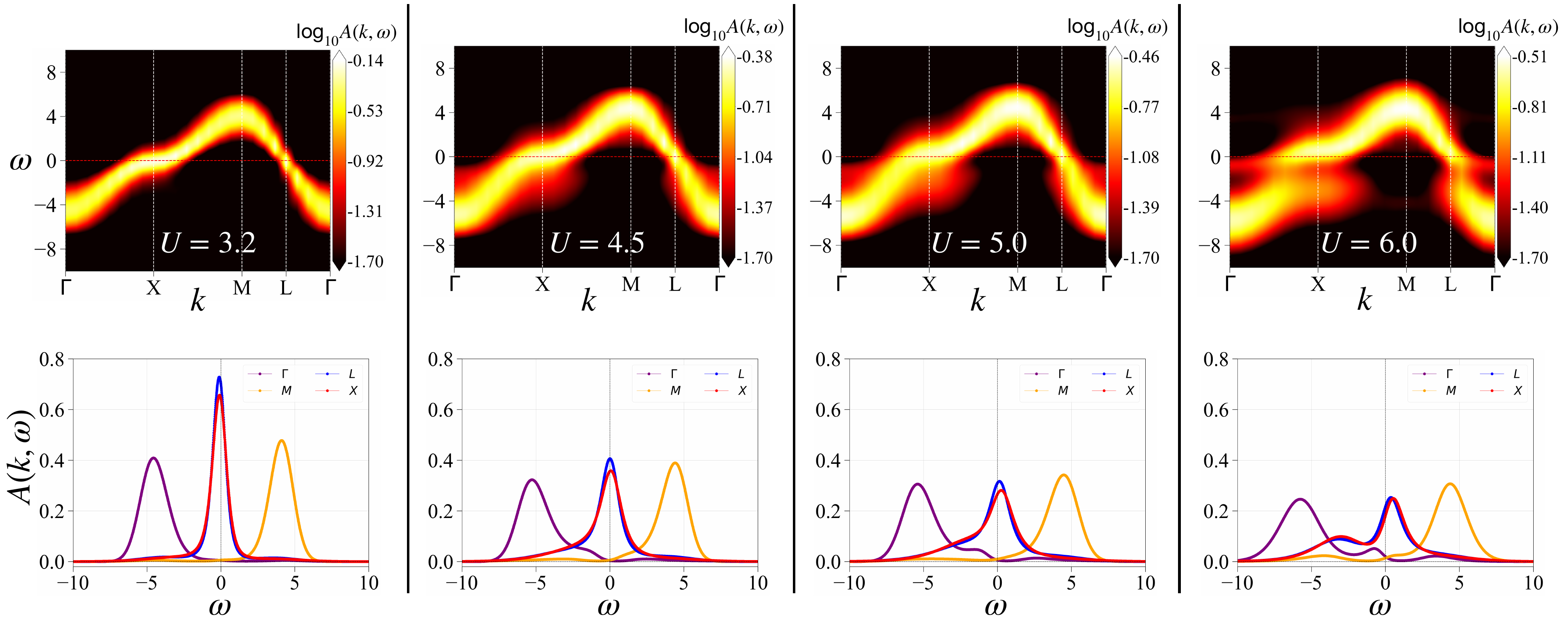}}
\caption{1-particle dispersions at fixed density $n=1.05$, at $T=0.5$. For $U<3.5$, the system is a metal at half filling; doping results in a metallic state with ``quasiparticle" like sharp excitation peak around the Fermi level $\omega=0$. For $3.5 \leq U \leq 5$, the system is non-Fermi liquid at half filling. As the system is doped from the AM phase, a fraction of the spectral weight near $(0,\pi)$ point gets pushed down below $\omega=0$, leading to the development of shoulder at $\omega <0$ in $N(\omega)$, shown in main text. With increasing $U$, the spectral weight transfer below the Fermi level increases, leading to a prominent lower Hubbard band like feature in $N(\omega)$, as seen for doped Mott insulators.} 
\label{Figure2_end_matter}
\end{figure*}

The Fermi surface reconstruction inferred from the sign change of $\text{Re}G(k,\omega=0)$ is independently reflected in the momentum distribution curves $A(k,\omega=0)$, shown in Fig.~\ref{Figure1_end_matter}(b). For $U=3.2$, corresponding to doping from the metallic regime at half filling, the spectral intensity $A(k,\omega=0)$ exhibits pronounced peaks at the locations expected from the non-interacting Fermi surface. As the interaction strength is increased into the AM regime ($U=4.0,4.5,4.8$), the low-energy spectral weight near the antinodal $X$ point is progressively suppressed and shifted away from the non-interacting Fermi-surface location, while the spectral weight near the nodal $L$ point remains comparatively robust. This momentum-selective suppression directly mirrors the spectral-weight redistribution discussed in Fig.~\ref{Figure2_end_matter}.

The largest deviations from the non-interacting Fermi surface occur near the $X$ point and indicates that the precursor Mottness does not affect all momenta equally. Instead, charge localization first suppresses low-energy excitations in specific regions of the Brillouin zone while leaving other portions of the Fermi surface relatively intact. This can be attributed to the fact that the single particle dispersions $A(k,\omega)$ are ``flatter" near the $X$ point compared to the $L$ point, and hence the effect of interactions are stronger here.

\section{Single-particle spectra for the doped anomalous metal}
\label{Spectral weight}
\noindent
In the main text, we showed how particle doping the anomalous-metallic phase at half-filling leads to an anomalous shoulder like feature for $\omega<0$ in the single-particle density of states $N(\omega)$, that evolves into a well separated lower Hubbard band for $U>U_{\rm cr}^{\rm DOS}(T)$. Here, we show the detailed single-particle spectra $A(k,\omega)$, when the system is doped from metallic, anomalous-metallic and insulating regime at half-filling.

The top panel of Fig.~\ref{Figure2_end_matter} shows the single-particle dispersion $A(k,
\omega)$ along a high symmetry path on a log scale (to resolve finer features), and the bottom panel shows cuts of spectral functions at the high symmetry points. When the system is doped from the metallic regime at half-filling ($U=3.2$), $A(k,\omega)$ shows a single coherent dispersion, and the spectral functions at the high symmetry points show a single peak structure, with the $L$ and $X$ points showing a significantly higher weight than $X$ and $M$ points. As the system is doped from the anomalous-metallic regime, $(U=4.5)$, spectral weight from the Fermi level shift down below it, which shows up as a smearing of the dispersion, most dominantly around the $X$ points. The spectral function cuts indeed confirm this. An asymmetric shoulder starts to develop for $\omega<0$ for the $X$ and $L$ points; in addition, the spectral function at $\Gamma$ also start to develop a bump near $\omega=0$. This leads to the development of the shoulder seen in $N(\omega)$ for $\omega<0$ in the main text. Also note that the peak of $A(k,\omega)$ at $X$ and $M$ points decrease in magnitude compared to $\Gamma$ and $M$ points; this is analogous to the spectral weight redistribution mechanism at half-filling, discussed above and also in the main text.

As the system crossovers to the insulating regime at half-filling, $U=5.0$, doping it enhances this spectral anomaly. The shoulder at the $X$ and $M$ points, as well as the low energy bump $\Gamma$ point get more pronounced, with the magnitude of $A(k,\omega)$ near the Fermi level for $X$ and $L$ points now lower than the peak of $A(k,\omega)$ for $\Gamma$ and $M$ points. Finally, at larger values of $U=6.0$, the spectra shows separation into three main branches in the top panel. The spectral cuts in the bottom panel indeed confirms this: The spectral function $A(k,\omega)$ at the $X$ and $L$ points now feature a prominent low energy peak at $\omega<0$, and the spectra at $\Gamma$ point evolves into having a small bump near $\omega=0$. The separation of a single peak at the $X$ and $M$ points into two peaks results in a well separated lower Hubbard band in $N(\omega)$. Although the anomalous metallic regime at half-filling shows no incipient lower and upper Hubbard bands, doping it results in an asymmetric spectra, that already starts to show development of lower Hubbard band like signatures. Thus, the observation of a three peak like structure in $N(\omega)$ for the doped system, characteristic of doped Mott Insulator, doesn't require the presence of a gapped spectra at half-filling. 
\typeout{}

%\todo{Add ReG(k,0) data for U=2.0,3.2,3.5,4.0,4.5,4.6,4.7,4.8,5.0,5.5,6.0,8.0 in appendix. Also spectral functions at $(0,\pi)$ point.}
%\todo{For half filling, Fig 10, replace gaussian maxent plots with that from uniform default model}
%\todo{Also add the plots for Im Sigma at first matsubara frequency, as well as their slope along a high symmetry path. Contrast with Re G and A(k,w)}
%\todo{Provide the $\chi_{q,\tau}$ data in the appendix to prove that the error bars are all well behaved}

\bibliography{ref}

\clearpage
\onecolumngrid

\begin{center}
{\Large Supplemental Material: Finite temperature precursors of Mottness in the Fermi Hubbard model}
\end{center}

\vspace{2em}

\appendix

% Optional: reset numbering
\setcounter{section}{0}
\setcounter{figure}{0}
\setcounter{table}{0}
\setcounter{equation}{0}

\renewcommand{\thesection}{S\Roman{section}}
\renewcommand{\thefigure}{S\arabic{figure}}
\renewcommand{\thetable}{S\arabic{table}}
\renewcommand{\theequation}{S\arabic{equation}}
\twocolumngrid

\section{Vertex contribution to two-particle responses}

The two particle response functions are defined as:

\begin{align}
    &\chi(q,i\omega_n) = \frac{-1}{\beta}\sum_{k,i\nu,\sigma}v_{k\sigma} G(k,i\nu)G(k+q,i\nu+i\omega_n)v_{k+q,\sigma} \nonumber \\
    &-\frac{1}{\beta^2}\sum_{k,i\nu,\sigma}\sum_{k',i\lambda,\sigma'}v_{k\sigma}G(k,i\nu)G(k+q,i\nu+i\omega_n) \nonumber \\
    &\times \Gamma^{kk'q}_{\sigma \sigma'}(i\nu,i\lambda;i\omega_n)G(k',i\lambda)G(k'+q,i\lambda+i\omega_n)v_{k'+q,\sigma'}.. 
\label{EqA1_two_particle_response}
\end{align}

where the object $\Gamma^{kk'q}_{\sigma \sigma'}(i\nu,i\lambda,i\omega_n)$ is the fully irreducible vertex in the particle hole channel~\cite{georges1996dynamical}. The spin dependence of the Green's function has been absorbed into the response factors $v_{k,\sigma}$, defined as~\cite{georges1996dynamical,schafer2016nonperturbative}

\begin{align}
    \chi_{\rm ch}(q,i\omega_n): ~v_{k,\sigma} = 1,~~\chi_{\rm spin}(q,i\omega_n): ~v_{k,\sigma} = sgn(\sigma) %\nonumber \\
    %\chi^{xx}_{curr}(q,i\omega_n)&: ~v_{k,\sigma} = -2te\sin(k_x)
\end{align}

The electronic compressibility is defined as $\tilde{\kappa} = \frac{\partial n}{\partial \mu}$. We wish to show the following identity:

\begin{align}
    \frac{\partial n}{\partial \mu} = \chi_{\rm ch}(q=0,\omega=0) = \frac{-1}{\pi}\int_{-\infty}^{\infty}d\omega \frac{\chi''_{\text{ch}}(q=0,\omega)}{\omega}
\end{align}

Here, $\chi''$ is the imaginary part of the two particle response function. This is proved by switching to the Lehman representation;

\begin{align}
    \chi_{\text{ch}}(q,\omega) = \frac{1}{Z}\sum_{n,m}\frac{(e^{-\beta E_n}-e^{-\beta E_m})}{\omega+i\eta +E_n-E_m}\langle n |\rho_q|m\rangle \langle m|\rho_{-q}|n\rangle
\end{align}

where the density operator is defined as $\rho_{q} =\sum_{k,\sigma}c^{\dagger}_{k+q,\sigma}c_{k,\sigma}$. Taking the $q\rightarrow 0$ limit of this yields the number operator for the total density, $\hat{N} = \sum_{k,\sigma}c^{\dagger}_{k,\sigma}c_{k,\sigma}$. 

\begin{align}
    &\frac{\chi''(q\rightarrow 0,\omega)}{\omega} = -\pi\sum_{n,m}\frac{(e^{-\beta E_n}-e^{-\beta E_m})}{\omega} \nonumber \\
    &~~~~~~~~~~~~~~~~~~~~~~~~~~~\times \langle n|\hat{N}|m\rangle \langle m|\hat{N}|m\rangle\delta (\omega+E_n-E_m) \end{align}
The imaginary part of the response function thus becomes~\cite{peterson2010kelvin}
    \begin{align}
    & \frac{-1}{\pi}\int d\omega \frac{\chi''(q\rightarrow 0,\omega)}{\omega} = \sum_{n,m}\frac{(e^{-\beta E_n}-e^{-\beta E_m})}{E_n-E_m} \nonumber \\
    &\times\langle n|\hat{N}|m\rangle \langle m|\hat{N}|m\rangle = V\frac{\partial n}{\partial \mu}
\end{align}

where $V$ is the volume of the system. Thus, the compressibility can be calculated either by taking the $i\omega_n=0$ component of $\chi_{\rm ch}$ in Matsubara frequency, or by looking at the frequency integrated slope of the spectral part of $\chi_{\rm 
ch}$ in real frequency. 

Next, we split the total contribution to the compressibility into two sources, the bubble contribution (in the particle hole channel), and the vertex contribution

\begin{align}
    \frac{\partial n}{\partial \mu} = \frac{-1}{\pi}\int d\omega \frac{\chi''_{\text{bubble}}(0,\omega)}{\omega}+\frac{-1}{\pi}\int d\omega \frac{\chi''_{\text{vertex}}(0,\omega)}{\omega}
    \label{EqA7: bubble vertex}
\end{align}

For the two particle response in Eq~\ref{EqA1_two_particle_response}, the bubble term is the polarization bubble (suppressing the response factors $v_{k\sigma}$ in Eq~\ref{EqA1_two_particle_response} for now),

\begin{align}
    \chi_{\text{bubble}}(q,i \omega_n) = -\frac{1}{\beta}\sum_{k,i\nu,\sigma}G(k,i\nu)G(k+q,i\omega_n+i\nu)
    \label{EqA8: Bubble term matsubara freq}
\end{align}
Note that in QMC, since we directly work with finite temperature data, it is favorable to compute the bubble contribution by directly computing the Matsubara sum, Eq~\ref{EqA8: Bubble term matsubara freq}. %Owing to the definition of the Matsubara Green's function,

%\begin{align}
%    G(k,i\nu) = \int d\omega \frac{A(k,\omega)}{i\nu-\omega}
%\end{align}

%The Green's function satisfy the identity, $G(k,-i\nu) = (G(k,i\nu))^{*}$. Hence, when we compute the static, uniform part of the bubble, $\chi_{\rm bubble}(q=0, i\omega_n=0)$, it is sufficient to restrict the sum over only positive Matsubara frequencies to get

%\begin{align}
%   \chi_{\rm bubble}(0,0) = \frac{2}{\beta}\sum_{k,\nu>0}2[\text{Re}G(k,i\nu)]^2 +2[\text{Im}G(k,i\nu)]^2 
%\end{align}

%This guarantees that the bubble contribution to the compressibility is always real. 
The Matsubara sum can also be performed using standard techniques to express results in real frequency $\nu$; we get (replacing $i\omega_n \rightarrow z$, with Im$z>0$);

\begin{align}
    \chi_{\text{bubble}}(q,z) &= \frac{1}{\pi}\sum_{k,\sigma}\int_{-\infty}^{\infty}d\nu n_{f}(\nu)\big [Im G(k,\nu)G(k+q,z+\nu) \nonumber \\
    &+G(k,\nu-z)Im G(k+q,\nu) \big] \nonumber \\
    &=\sum_{k,\sigma}\int_{-\infty}^{\infty}d\nu n_{f}(\nu)\big [A(k,\nu)G(k+q,z+\nu) \nonumber \\
    &+G(k,\nu-z)A(k+q,\nu) \big]
\end{align}

where $n_{f}(\nu) = \frac{1}{1+e^{\beta \nu}}$ is the fermi distribution function at the inverse temperature $\beta$. The imaginary part of the response can be obtained by $\chi^{''}(q,z) = \frac{1}{2i}[\chi(q,z^{+})-\chi(q,z^{-})]$. Using this, we get

\begin{align}
    &\chi^{''}_{\text{bubble}}(q,z) = \sum_{k,\sigma} \int _{-\infty}^{\infty}d\nu n_{f}(\nu)\frac{-1}{2i}[A(k,\nu) \nonumber \\
    &\{G(k+q,z^{+}+\nu) -G(k+q,\nu+z^{-})\} \nonumber \\
    &+A(k+q,\nu)\{G(k,\nu-z^{+})-G(k,\nu-z^{-}\}]
\end{align}

Analytically continuing to real frequency $z\rightarrow \omega+i\eta$, and putting in the response factors $v_{k\sigma}$ back, we get

\begin{align}
    &\chi_{\text{bubble}}^{''}(q,\omega) = -\pi \sum_{k,\sigma}v_{k,\sigma}v_{k+q,\sigma}\int_{-\infty}^{\infty}d\nu n_{f}(\omega)\{A(k,\nu)\nonumber \\
    &\times A(k+q,\nu+\omega)-A(k+q,\nu)A(k,\nu-\omega)\} \nonumber \\
    &=-\pi\sum_{k,\sigma}v_{k,\sigma}v_{k+q,\sigma}\int_{-\infty}^\infty d\nu [n_{f}(\nu)-n_{f}(\nu+\omega)] \nonumber \\
    &~~~~~~~~~~~~~~~~~~~~~~~~~\times A(k,\omega)A(k+q,\nu+\omega)
    \label{EqA11: Bubble real freq}
\end{align}

The bubble contribution to the compressibility is thus given by (using Eq~\ref{EqA7: bubble vertex}): 

\begin{align}
    \frac{\partial n}{\partial \mu}= \frac{2}{V}\sum_{k}\int d\omega d\nu \frac{[n_{f}(\nu)-n_{f}(\nu+\omega)]}{\omega}A(k,\omega)A(k,\omega+\nu)
\end{align}
Note that the above Eq~\ref{EqA11: Bubble real freq} serves as an independent check of our analytic continuation. The bubble contribution computed from the Matsubara sum in Eq~\ref{EqA8: Bubble term matsubara freq} coincides with the bubble contribution calculated from the analytically continued real frequency Green's functions, within error bars.

\begin{figure}[t]{\includegraphics[width=0.8\linewidth]{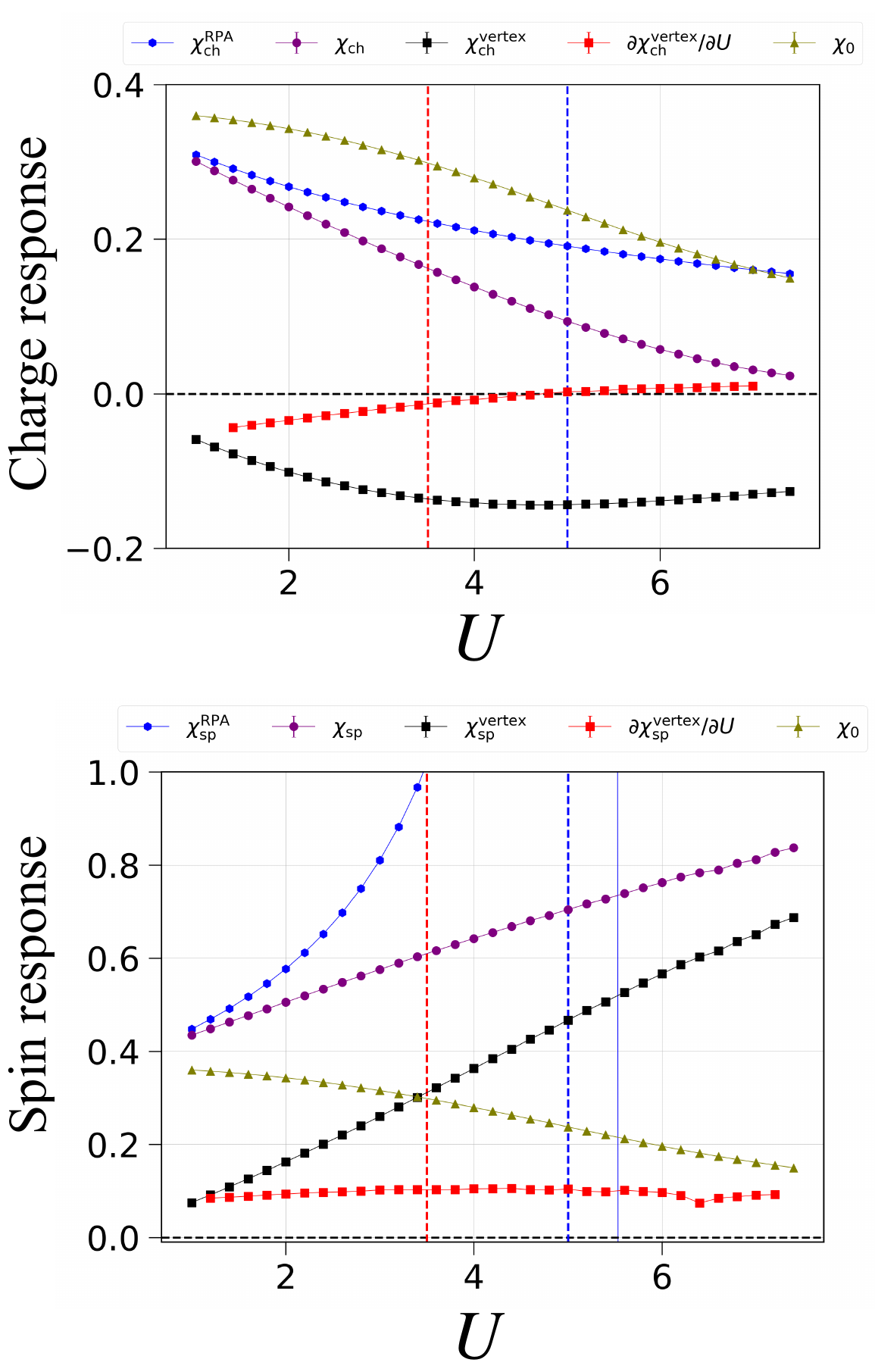}}
\caption{Static uniform charge and spin response across the extended crossover. \textbf{Top panel}: Static charge response. \textbf{Bottom panel}: Static spin response. The purple circles show the full response ($\chi_{\rm ch} = \frac{\partial n}{\partial \mu}$, $\chi_{\rm sp} = \frac{\partial m}{\partial h}$), while the blue circles denote the corresponding RPA response. The black squares denote the vertex contribution. In the charge channel, the vertex contribution is negative and suppresses charge fluctuations, with its magnitude reaching a maximum in the vicinity of the anomalous-metal regime before decreasing toward the insulating state. In contrast, the spin-channel vertex contribution is positive and grows monotonically with interaction strength, enhancing magnetic fluctuations. The vertical red and blue dashed lines mark the onset of the anomalous-metallic regime, $U_{\rm cr}^{\rm TDOS}(T)$, and the onset of the insulating regime, $U_{\rm cr}^{\rm DOS}(T)$, respectively.} 
\label{Figure1_appendix}
\end{figure}

We can isolate the bubble and vertex contributions to the spin and charge response, this is shown in Fig~\ref{Figure1_appendix}. We define the vertex contribution as

\begin{align}
\chi^{\rm vertex}_{\rm ch/sp} = \chi_{\rm ch/sp}-\chi^{\rm bubble}_{\rm ch/sp}
\end{align}

where we have isolated the bubble contribution as the first term on RHS of Eq~\ref{EqA1_two_particle_response}, and the vertex contribution as the remainder of the term of the infinite series. The bubble contribution is given by Eq~\ref{EqA8: Bubble term matsubara freq}. The full charge response is given by $\chi_{\rm ch} = \frac{\partial n}{\partial \mu}$, and the spin response is given by $\chi_{\rm sp} = \frac{\partial m}{\partial h}$. For the latter, estimating the spin response by applying a magnetic field and taking the numerical derivative of the magnetization gives noisier results. Instead, we extract it from the fluctuation dissipation theorem, $\chi_{\rm sp} = \beta\sum_{\langle ij\rangle}\langle \hat{S}^{z}_{i}\hat{S}^{z}_{j}\rangle_{c}$, and we define the spin operators $\hat{S}^{z}_{i} = (n_{i\uparrow}-n_{i\downarrow})$. The latter quantity is calculated on several large sized lattice ($N=16\times 16$ and $N=20\times 20$) to ensure finite size effects are negligible. The RPA response functions are given by

\begin{align}
    \chi^{\rm RPA}_{\rm ch} = \frac{\chi^{\rm bubble}_{0}}{1+\frac{U}{2}\chi^{\rm bubble}_{0}} ,~~~~\chi^{\rm RPA}_{\rm sp} = \frac{\chi^{\rm bubble}_{0}}{1-\frac{U}{2}\chi^{\rm bubble}_{0}}
\end{align}

where $\chi^{\rm bubble}_{0} = N_{0}(0)$ is the bubble term with the non-interacting propagators, and equals the density of states of the non-interacting system at the same temperature at chemical potential.

\begin{figure*}{\includegraphics[width=\linewidth]{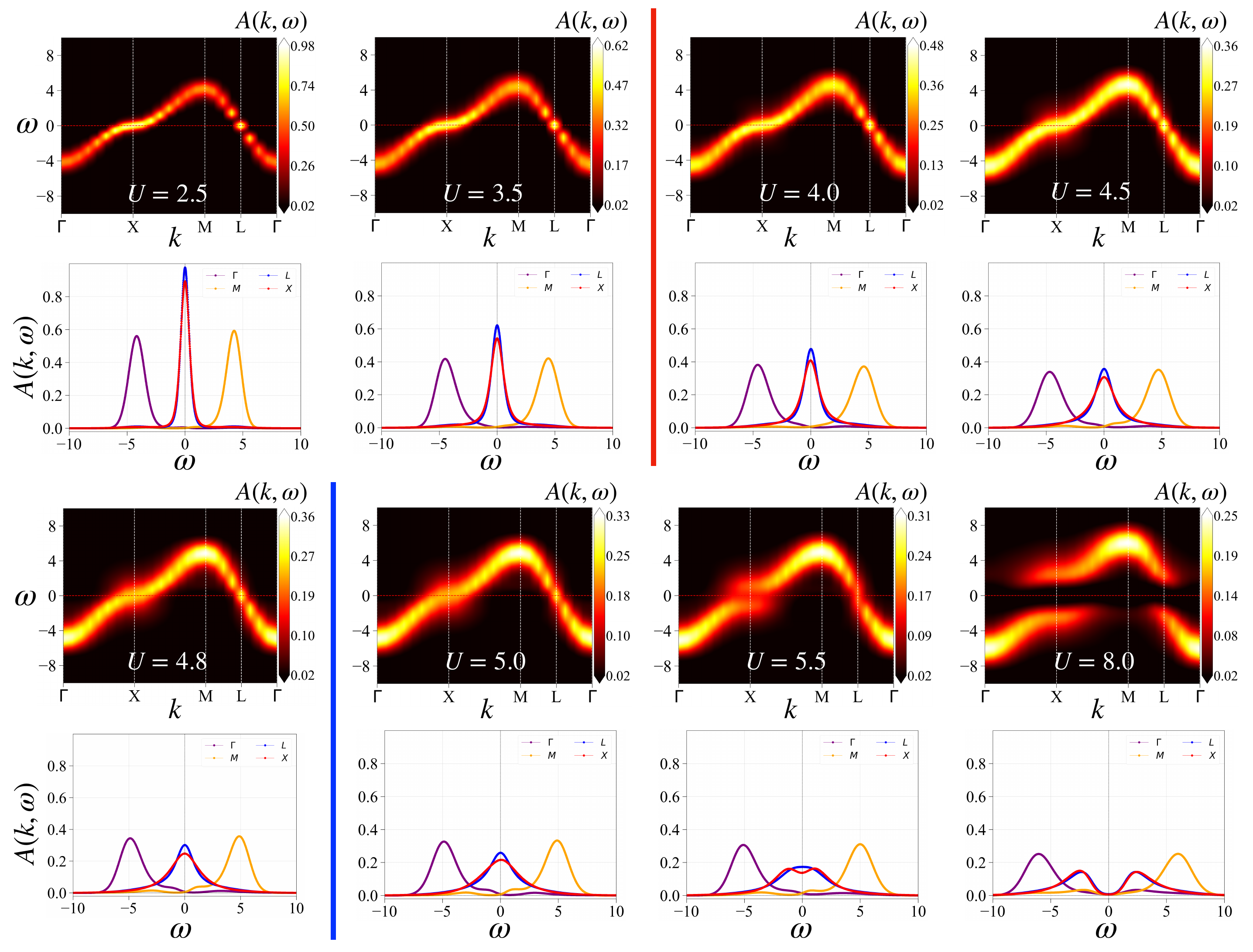}}
\caption{Evolution of the momentum-resolved spectral function $A(k,\omega)$ across the extended crossover at half filling and $T=0.5$. The upper panels show the dispersion along the high-symmetry path $\Gamma\rightarrow X\rightarrow M\rightarrow L\rightarrow \Gamma$, while the lower panels show cuts of the spectra at the $\Gamma$, $X$, $M$, and $L$ points. The red vertical line marks the onset of the anomalous-metallic (AM) regime at $U_{\rm cr}^{\rm TDOS}(T)$, while the blue vertical line marks the onset of the insulating regime at $U_{\rm cr}^{\rm DOS}(T)$. Across the crossover, the low-energy spectral weight near the $X=(0,\pi)$ and $L=(\pi/2,\pi/2)$ points is progressively reduced, accompanied by enhanced spectral weight near the $\Gamma$ and $M$ points. Around $U\sim5.0$, the redistribution becomes sufficiently strong to produce a dip in the single-particle density of states $N(\omega)$, even though the momentum-resolved spectra remain gapless. At larger interaction strengths, the suppression of low-energy spectral weight becomes stronger at the antinodal $X=(0,\pi)$ point than at the nodal $L=(\pi/2,\pi/2)$ point. For $U=8.0$, the spectra become fully gapped and evolve into well-separated upper and lower Hubbard bands.}
\label{Figure2_Appendix}
\end{figure*}

The RPA susceptibilities reproduce the qualitative evolution of the charge and spin responses across the crossover, namely the suppression of charge fluctuations and the enhancement of spin fluctuations with increasing $U$. Quantitatively, however, RPA substantially overestimates the spin response while remaining comparatively close to the charge response, demonstrating that interaction effects beyond the weak-coupling ladder resummation are considerably more important in the spin channel. As expected, the vertex contribution to the charge response is negative, since charge fluctuations are suppressed uniformly as $U$ in increased. In contrast, the vertex contribution to the spin response is always positive. However, note that while $\chi_{\rm sp}^{\rm vertex}$ grows monotonically with $U$, $\chi_{\rm ch}^{\rm vertex}$ is nonmonotonic in $U$. It grows in magnitude till $U_{\rm cr}^{\rm DOS}(T)$, and starts to decrease slowly once a gap at $\omega=0$ onsets in $N(\omega)$. At weak coupling, increasing $U$ suppresses charge fluctuations through the reduction of double occupancy, leading to increasingly strong vertex corrections. This suppression continues through the anomalous-metallic regime. However, as $N(\omega)$ develops an insulating character at $U_{\rm cr}^{\rm TDOS}(T)$,  low-energy charge fluctuations become strongly depleted, and the magnitude of the vertex contribution starts to decrease. In contrast, the spin vertex contribution remains positive and increases steadily with $U$. As charge fluctuations are suppressed, local moments become better defined, leading to an increasingly large enhancement of the spin response beyond the bubble contribution. The peak in the charge vertex contribution near the crossover therefore reflects the scenario where interaction effects most strongly suppress charge fluctuations, while the monotonic growth of the spin vertex contribution tracks the gradual development of local-moment physics.

%\begin{figure}{\includegraphics[width=\linewidth]{Graphs_new/Figure_appendix_6_self_energy.pdf}}
%\captionof{figure}{} 
%\label{Figure3_Appendix}
%\end{figure}

%\begin{figure}{\includegraphics[width=\linewidth]{Graphs_new/Figure_appendix_7_scattering_rate.pdf}}
%\captionof{figure}{} 
%\label{Figure3_Appendix}
%\end{figure}

\begin{figure}{\includegraphics[width=0.8\linewidth]{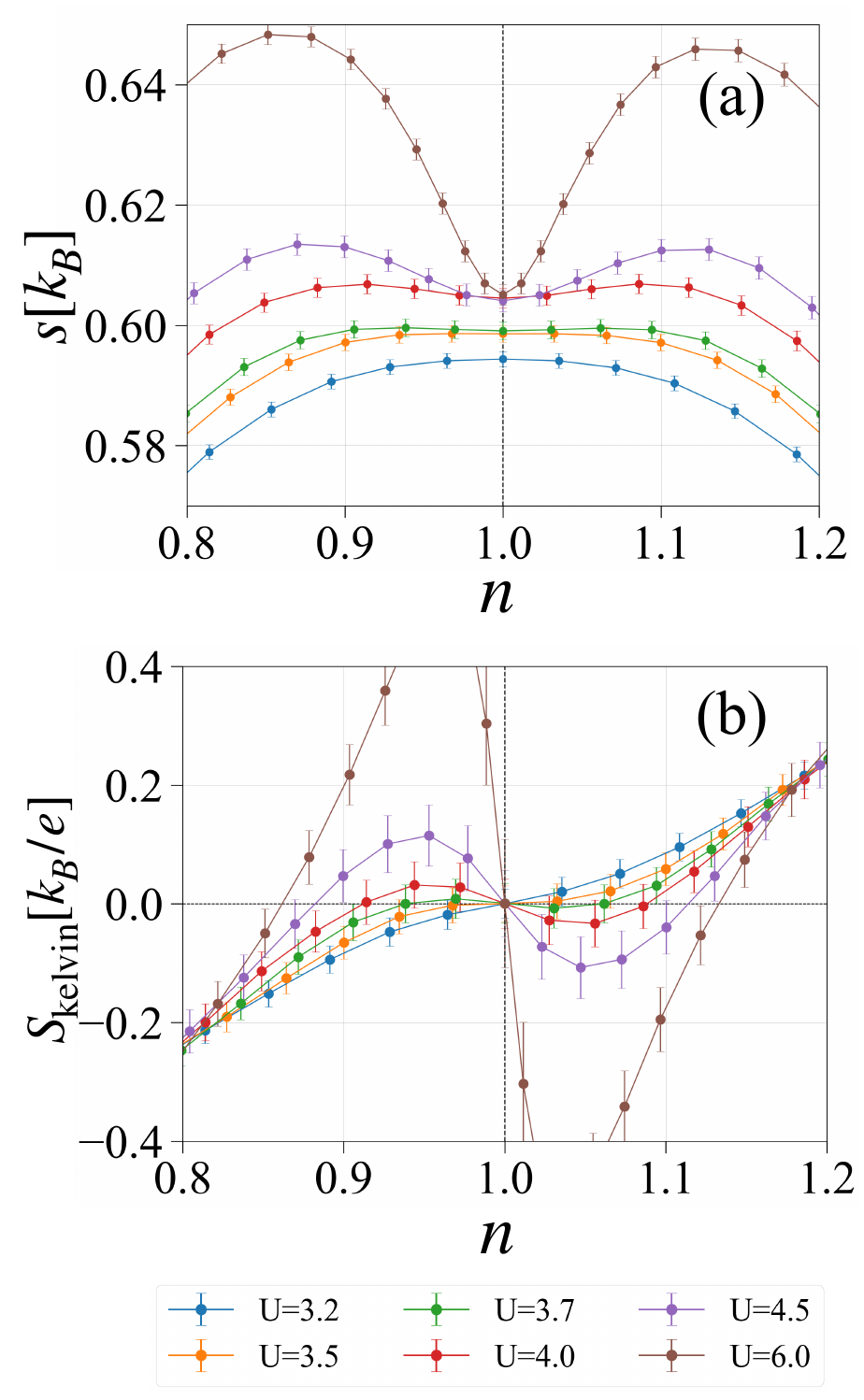}}
\caption{Entropy and Seebeck coefficient in the repulsive Hubbard model across the extended metal-insulator crossover at $T=0.5$. In the metallic regime, the entropy maxima sits at half filling. However, as the system is doped from half filling starting from the AM regime, the entropy maxima shifts to finite doping. \textbf{(b)} Seebeck coefficient determined from Kelvin formula. In the metallic phase, $U<3.2$, the sign of the Seebeck coefficient follows that of a non interacting system. However, as the system is doped from the AM regime at half filling, the Seebeck coefficient exhibits the ``wrong" sign, which signals a change in carrier type and is consistent with the anomalous Fermi surface in the doped anomalous-metallic case in the main text} 
\label{Figure4_Appendix}
\end{figure}
\section{Spectral weight redistribution across the extended crossover}

We look at the single-particle spectral function $A(k,\omega)$ at half filling, across the extended crossover discussed in main text. We consider a high symmetry path that goes from $\Gamma = (0,0) \rightarrow X (0,\pi) \rightarrow M(\pi,\pi) \rightarrow L(\pi/2,\pi/2) \rightarrow \Gamma (0,0)$. Note that because of the sum rule, $\frac{1}{N}\sum_{k}\int d\omega A(k,\omega)=1$, the total spectral weight in the entire Brilluoin zone is conserved. Hence, a loss of spectral weight at a certain $(k,\omega)$ should be compensated by a gain of spectral weight at $(k',\omega')$. We take this as the basis for discussing spectral weight redistribution. We show both the single-particle dispersion in the $(k,\omega)$ plane, as well as cuts along fixed $k$ of spectral functions, for several high symmetry points in Fig~\ref{Figure2_Appendix}.

For $U<U_{\rm cr}^{\rm TDOS}(T)$, the half filled system is in the metallic regime ($U=2.5,3.2$). The low-energy spectral weight is concentrated predominantly near the $X=(0,\pi)$ and $L=(\pi/2,\pi/2)$ points, which lie on the non-interacting Fermi surface. The corresponding spectral functions exhibit sharp peaks centered at $\omega=0$, indicating coherent low-energy excitations. 
Once the system crossovers into the anomalous-metallic (AM) regime ($U> U_{\rm cr}^{\rm TDOS}(T)$), the spectral weight near the $X$ and $L$ points begins to decrease continuously. Simultaneously, spectral weight accumulates near the $\Gamma$ and $M$ points. This redistribution is visible both in the momentum-resolved dispersions and in the individual cuts at the high-symmetry points. Importantly, throughout the AM regime the spectral functions at the $X$ and $L$ points still retain finite low-energy weight at $\omega=0$, even though the charge response has already become exponentially suppressed, as discussed in the main text. Thus, the onset of precursor Mottness first manifests through momentum-space redistribution and suppression of charge fluctuations, rather than through immediate gap formation in the momentum-resolved spectra.

The redistribution becomes significantly stronger upon approaching $U_{\rm cr}^{\rm DOS}(T)$. Around $U\sim5.0$, the peak of the spectral functions at $X$ and $L$ points are smaller than the peaks at $\Gamma$ and $M$ points; this leads to the gap onset in $N(\omega)$. This demonstrates that the onset of insulating behavior in the DOS at $U_{\rm cr}^{\rm DOS}(T)$ is driven by a redistribution of spectral weight in the Brillouin zone, rather than gap formation in spectral functions at individual momenta. Upon increasing $U$ further ($U=5.5$), $X=(0,\pi)$ point develops a pronounced suppression of low-energy spectral weight. In contrast, the nodal $L=(\pi/2,\pi/2)$ point still retains finite low-energy spectral weight. This momentum-selective suppression reflects the nodal–antinodal dichotomy of the Hubbard model: interactions are more pronounced at the $(0,\pi)$ point, compared to the $(\pi/2,\pi/2)$, due to a flatter dispersion near the $(0,\pi)$ point; this amplifies the effects of interactions. This is consistent with recent DQMC studies by Lu \textit{et al.}~\cite{lu2026quantum}, which also identified stronger pseudogap formation near the antinodal region across the finite-temperature crossover. At even larger $U=8.0$, the spectra becomes fully gapped, and the lower and upper Hubbard bands in $N(\omega)$ are now well separated, leading to the Mott Insulator. In the $(k,\omega)$ plane, these arise as two disconnected branches of the single-particle dispersion.

Our discussions provides a microscopic picture of the extended crossover discussed in the main text. At finite temperature, the onset of insulating behavior in $N(\omega)$ is realized not through simultaneous gap opening at all momenta, but through a progressive redistribution of spectral weight away from the Fermi-surface momenta. The antinodal region is affected first, followed by the nodal region at larger interaction strengths, ultimately leading to the formation of the large-$U$ Mott insulating state.

\section{Entropy and Seebeck coefficient}
The Kelvin formula allows us to relate the Seebeck coefficient to the thermodynamic entropy (defined per unit area)~\cite{peterson2010kelvin}, $s = \frac{1}{T}(\epsilon_k+\epsilon_p-\mu n+P)$, where $\epsilon_k, \epsilon_p$ are the kinetic and potential energy densities respectively, and $n$ is the number density. From the kelvin formula, the Seebeck coefficient is the slope of the entropy curve as a function of number density at fixed temperature, $S_{\rm kelvin} = -(\partial s/\partial n)_{T}$. In the non-interacting limit, the entropy is maximum at half filling, leading to a Seebeck coefficient that is positive for electron doping ($n>1$), negative for hole doping ($n<1$), and zero at half filling, ($n=1$) due to particle hole symmetry of the model. However, for stronger interactions $U>U_c(T)$, the maximum in the entropy shifts to a finite doping~\cite{lenihan2021entropy}. This can be seen by noting the Maxwell relation, 

\begin{align}
    \frac{\partial s}{\partial \mu}\bigg|_{T} = \frac{\partial n}{\partial T}\bigg|_{\mu} \implies \frac{\partial ^2 s}{\partial \mu^2} = \frac{\partial}{\partial T}\bigg(\frac{\partial n}{\partial \mu}\bigg) = \frac{\partial \tilde{\kappa}}{\partial T} 
\end{align}

For an entropy maxima at half filling, $\frac{\partial ^2 s}{\partial \mu^2}<0$ implies $\frac{\partial \tilde{\kappa}}{\partial T}<0$, which occurs for the metallic regime at half filling in Fig 1(a) of the main text. However, as the system crosses into the anomalous-metallic regime at $U_{\rm cr}^{\rm TDOS}(T)$, a gap onset in compressibility (TDOS) causes $\frac{\partial \tilde{\kappa}}{\partial T}>0$, and hence the entropy maxima at half filling changes into a minima, and the maxima occurs now at a finite doping. This causes the Seebeck coefficient to change sign at finite doping, shown in Fig~\ref{Figure4_Appendix}. 

The thermodynamic connection of the Seebeck coefficient with the compressibility allowed us to construct the phase diagram in Figure 1 of the main text. Thus, given a particular value of $(U,T)$, we can compute the temperature dependence of TDOS $\frac{\partial \tilde{\kappa}}{\partial T}$ at half filling, and determine if there will be an anomalous sign change of the Seebeck coefficient at finite doping. In Fig~\ref{Figure4_Appendix}(a-b), this is shown to be case at $T=0.5$. The location of the sign change corresponds to the ``isobestic" points in equation of states, where the equation of state around a particular chemical potential $\mu_{\rm ib}$ shows little variation in temperature, $(\partial n/\partial T)_{\mu_{\rm ib}}=0$

%\section{Crossover at lower temperatures}

\end{document}